\documentclass{article}[12pt]

\begin{document}

\title{\bf A Formula for Multi-Loop 4-Particle Amplitude in
Superstring Theory}

\author{Chuan-Jie Zhu\footnote{E-mail: zhucj@itp.ac.cn}
\thanks{Supported
in part by fund from the National Natural Science Foundation of
China with grant Number 10475104.} \\[3mm]\\
Institute of Theoretical Physics,
Chinese Academy of Sciences\\
P. O. Box 2735, Beijing 100080,
P. R. China\footnote{Permanent address}\\[3mm]\\
and   \\[3mm]\\
 Abdus Salam International Center for Theoretical Physics\\
Strada Costiera 11, I-34014 Trieste, Italy}

\maketitle

\abstract{Based on the recent developments of explicit
computations at 2 loops in superstring theory in the covariant RNS
formalism, we propose an explicit formula for the arbitrary loop
4-particle amplitude in superstring theory. We prove that this
formula passes two very difficult tests: modular invariance and
factorization. If proved, this shows that superstring theory is
not only finite order by order in perturbation theory but is also
exceptionally simple.}

\newpage

\section{Introduction}

String perturbation theory (for a general review see
\cite{DHokerPhongReview}) was a very active research area after
the first superstring revolution in 1984/85
\cite{GreenSchwarz,Witten,Gross}. The general belief is that
superstring theory gives a finite theory of quantum gravity order
by order in perturbation theory \cite{Green86}. Although we now
understood this in general terms \cite{PolchinskiBook}, the
details have never been clearly spelled out in full generality
\cite{Mandelstam,Taylor,Sen}. The main difficulty is due to the
problem of covariant quantization of superstring theory. The
Green-Schwarz superstring \cite{GreenSchwarz83} is a highly
nonlinear theory. Only by choosing the light-cone gauge  we have a
free theory on the world-sheet. It then becomes quite easy to
compute tree- and one-loop amplitudes \cite{Schwarz82,GSW}. Due to
the choice of the non-covariant light-cone gauge (and also the
complicated dependence on interaction points) it is quite
difficult if not impossible to compute multi-particle and
higher-loop amplitudes.

The other approach is the covariant Ramond-Neveu-Schwarz (RNS)
formalism \cite{Knizhnik,FMS}. Here we can use the full power of
super-conformal field theory. The problem with this formalism is
that we must sum over all spin structures to get a consistent
and/or supersymmetric theory \cite{GSO,SeibergWitten}. Also the
naive integration over super-moduli space gives a result which
depends on the insertion points  of the super-Beltrami
differentials \cite{VerlindePLB}. So the results depend on these
spurious unphysical poles. Although these poles are harmless to
the physical amplitudes (they are total derivatives in moduli
space), they give complications to the analysis of the finiteness
property of the physical amplitudes.

At two loops the problem with modular invariance was solved by
Gava, Iengo and Sotkov in \cite{GavaIengoSotkov}. They  showed
that modular invariance at two loops fixed a unique way of
summation over spin structures and proved the vanishing of the
cosmological constant. Iengo and the author then used this result
to show the nonrenormalization theorem and to compute explicitly
the non-vanishing 4-particle in superstring theory
\cite{IengoZhu}. As expected the result (before integration over
moduli space) depends on the choice of the insertion points. It is
proved that changing the insertion points indeed give a total
derivative in moduli space. The apparent contradiction with the
$R^4$ conjecture \cite{GrossWitten, GreenGutpler} was solved by
recasting the amplitude into an explicitly modular invariant form
\cite{IengoZhu2} and the 2-loop correction to $R^4$ term was
indeed computed to be vanishing \cite{Iengo}.

Recently the problem of the arbitrariness at 2 loops was finally
solved by D'~Hoker and Phong in a series  of papers
\cite{DHokerPhongAll}. Basically D'~Hoker and Phong started from
first principles and gauge slice independence are kept at every
stage of the evaluation by using the method of projection onto
super period matrices. A gauge slice independent new measure was
obtained. Explicit computations were then carried out by using
this new measure and indeed a gauge slice independent result was
obtained \cite{ZhengWuZhuAll}. This result was also derived more
rigorously in \cite{DHokerPhongNow}. Recently D'~Hoker and Phong
\cite{DHokerPhong3loop} also gave a measure for three loop
superstring theory. It  remains to see if this can be used to do
explicit three loop computations in superstring theory.

Another promising approach of covariant quantization of
superstring is Berkovits' pure spinor approach \cite{Berkovitsa,
Berkovitsb,Berkovitsc} or closely related ones \cite{Othersa,
Othersb, Othersc, Othersd}. Recently this has been used for
proving the general non-renormalization theorem \cite{Berkovitsd}.
Also the leading contribution of the 4-particle amplitude is shown
to be 0, in agreement with the $R^4$ conjecture. It remains to see
how this method can be used to compute the full non-vanishing
4-particle amplitude.

Explicit result for higher loop amplitudes in superstring theory
is quite rare and so the known result should be exploited to
obtain as much information as possible. To our knowledge the only
explicitly known higher loop ($\ge 2$) non-vanishing amplitude is
the 2-loop four-particle amplitude in superstring theory, firstly
obtained in \cite{IengoZhu} and later re-obtained in
\cite{ZhengWuZhuAll,DHokerPhongNow} in an explicitly gauge
independent way. Although this result is somehow quite unique in
itself, it nevertheless  shows some surprising simplicity and
pattern which may be generalized straightforwardly to
higher-loops.  The first pattern is a complete cancellation of
determinant factors after summation over spin structures. The
second pattern is the appearance of holomorphic abelian
differentials in the integrand. This  appearance of holomorphic
abelian differentials  is also found in the explicit multi-loop
computations of Antoniadis, Gava, Narain and Taylor in
\cite{GavaNarain}. Although they computed a much simpler amplitude
which reduces to the topological string amplitude, we found the
appearance of the determinant of the abelian differentials
tantalizing. We will generalize these two patterns to multi-loops
and propose an explicit formula for the arbitrary loop 4-particle
amplitude in superstring theory. To our surprise, this formula can
pass two very difficult tests: modular invariance and
factorization. We hope that this formula will be proved someday.
If finally proved, this shows that superstring theory is not only
finite order by order in perturbation theory but is also
exceptionally simple. Generalization of the formula to heterotic
string theory may be easily found and it may have far reaching
consequence for the computations of higher loop amplitudes in the
maximal supersymmetric gauge theory \cite{Bern, Dixon}.

\section{The 4-particle amplitude in superstring theory}

Before we gave our formula for multi-loops, let us first recall
the 2-loop 4-particle amplitude obtained in
\cite{IengoZhu,ZhengWuZhuAll,DHokerPhongNow} is:
\begin{equation}
A_4^{\rm 2-loop} = K \, \int_{{\cal M}_g} { |\prod_{I\le J}^2 {\rm
d} \Omega_{IJ}|^2   \over (\det {\rm Im} \, \Omega)^5 }\,
\int_{\Sigma^4} |{\cal Y}_s|^2 \, {\rm
exp}\bigg(-\sum_{i<j}k_i\cdot k_j\,G(z_i,z_j)\bigg) ,
\label{known}
\end{equation}
where
\begin{eqnarray}
{\cal Y}_S &=& + (k_1-k_2)\cdot(k_3-k_4) \, \Delta (z_1,z_2)
\Delta (z_3,z_4)
\nonumber\\
& & + (k_2\leftrightarrow k_3, z_2\leftrightarrow z_3) +
 (k_2\leftrightarrow k_4, z_2\leftrightarrow z_4)  \nonumber \\
& = & (t-u) \Delta(z_1,z_2)\Delta(z_3,z_4) + (t-s)
\Delta(z_1,z_3)\nonumber \\
&& \times \Delta(z_2,z_4) + (u-s)
\Delta(z_1,z_4)\Delta(z_3,z_3) ,  \label{knowny} \\
\Delta(z,w) & = & \omega_1(z)\omega_2(w) - \omega_2(z)\omega_1(w),
\\
 G(z,w) & = & - \ln |E(z,w)|^2 + 2 \pi {\rm Im}\int_z^w \omega_I
({\rm Im}\Omega)^{-1}_{IJ}  {\rm Im} \int_z^w \omega_J .
\end{eqnarray}
As we said in the introduction, there is no determinant factors in
(\ref{known}). The integrand ${\cal Y}_s$ is defined only in terms
of the holomorphic abelian  differentials $\omega_i(z)$. In fact
this two patterns can be straightforwardly generalized to
arbitrary loops.

Our proposed formula for the $g$-loop 4-particle amplitude in
superstring theory is:
\begin{equation}
A_4^{g-{\rm loop}} = K \int_{{\cal M}_g} { |\wedge^a W^a \wedge_i
W_i|^2 \over (\det {\rm Im} \, \Omega)^5 }\, \int_{\Sigma^4}
|{\cal Y}_s|^2 \, {\rm exp}\bigg(-\sum_{i<j}k_i\cdot
k_j\,G(z_i,z_j)\bigg) . \label{conjecture}
\end{equation}
By comparing with the 2-loop expression given in eq.~(\ref{known})
we need to explain what is the measure $\wedge^a W^a \wedge_i
W_i$. ${\cal Y}_s$ is the same expression given in (\ref{knowny})
but with an appropriately  generalized $\Delta(z,w)$ (denoted as
$\Delta_g(z,w)$ to indicate it's for $g$-loop).

For a given Riemann surface, we fix an arbitrary generic
1-differential $\omega(z)$. This has $2g-2$ zeroes which we
denoted as $P_1, \cdots, P_{2g-3}$ and $P_{2g-2}=Q$. To construct
a basis of holomorphic 2-differentials, we also need a normalized
Abelian differential of the 3rd kind which is defined to have
vanishing $A$-period and to have simple poles at two different
points $P$ and $Q$ with residues $\pm 1$. Explicitly we have:
\begin{equation}
\omega_{PQ} (z) = d\, \ln(E(z,P)/E(z,Q)),
\end{equation}
where $E(z,P)$ is the prime form.

By using the given $\omega(z)$, we can construct a basis for the
holomorphic 2-differentials as follows \cite{IengoMarisa}:
\begin{eqnarray} \phi^a & = & \omega(z) \omega_{P_aQ}(z), \quad a =
1, \cdots, 2g -3,
\\
\phi_i & = & \omega(z) \omega_{i}(z), \quad i = 1, \cdots, g.
\end{eqnarray}
By the well-known correspondence between holomorphic
2-differentials and the holomorphic cotangent space of the moduli
space of Riemann surface, we have following elements in the
holomorphic cotangent space \cite{Manin}:
\begin{equation}
W^a = k(\phi^a), \qquad W_i = k(\phi_i).
\end{equation}
To refresh our memory we quote the well-known result:
\begin{equation}
k(\omega_i\omega_j)  = {1\over 2\pi i } {\rm d} \Omega_{ij} ,
\end{equation}
where $(\Omega_{ij})$ is the period matrix of the Riemann surface.
By taking the wedge products of all $W^a$ and $W_i$, we obtain a
volume form in moduli space which can be used to integrate over
the moduli space. This give the measure factor $\wedge^a W^a
\wedge^i W_i$ in eq.~(\ref{conjecture}).

The generalized $\Delta_g(z_i,z_j)$ is constructed also by making
use the zeroes of $\omega(z)$. Setting
\begin{equation}
\det\omega(z_1, z_2, \cdots, z_g) \equiv
\det( \omega_i(z_j)),
\end{equation}
we defined
\begin{eqnarray}
\Delta_g(z_i, z_j) & = & \sum_{\sigma'(1)=1} (-1)^{{\rm
sgn(\sigma)}} \det\omega(z_i, P_{\sigma(1)}, \cdots,
P_{\sigma(g-1)}) \nonumber
\\
& & \times  \det\omega(z_j, P_{\sigma(g)}, \cdots,
P_{\sigma(2g-2)})  - (z_i \leftrightarrow z_j) ,
\end{eqnarray}
where the summation over all permutation $\sigma$ is restricted to
$\sigma(1)=1$ and there is no summation over $\sigma$ which only
changes the ordering of $P$'s within the 2 determinants. In total
there are $2 \times {(2g-3)!/(g-1)!/(g-2)!}$ different terms.

For later use we can also use eq.~(\ref{conjecture}) to obtain the
g-loop 3-particle amplitude with one massive tensor  or 2-particle
self-energy correction to massive tensor, by factorizing 2
massless particles. The result is:
\begin{equation}
A_n^{g-{\rm loop}} = K_n \int_{{\cal M}_g} { |\wedge^a W^a
\wedge_i W_i|^2 \over (\det {\rm Im} \, \Omega)^5 }\,
\int_{\Sigma^n} |{\cal Y}_s^{(n)}|^2 \, {\rm
exp}\bigg(-\sum_{i<j}^nk_i\cdot k_j\,G(z_i,z_j)\bigg) ,
\label{conjecturea}
\end{equation}
where
\begin{eqnarray}
{\cal Y}_s^{(3)} & = & \Delta_g(z_1,z_3) \Delta_g(z_2,z_3), \\
{\cal Y}_s^{(2)} & = & (\Delta_g(z_1,z_2))^2 .
\end{eqnarray}

To finish this section we note that a direct application of the
proposed formula of (\ref{conjecture}) at two loops doesn't give
the known 2-loop result of (\ref{known}). In fact
$\Delta_g(z,w)=\Delta(z,w)\Delta(P_1,P_2)$ which vanishes
identically at two loops because $\Delta(P_1,P_2)=0$ when
$P_{1,2}$ are the zeroes of an abelian differential. This doesn't
happen for higher loops.  The 2-loop result can be obtained
correctly if we perturb the zero points $P_{1,2}$. This is in fact
necessary because of the special $Z_2$ symmetry at two loops which
is manifest by using the hyperelliptic representation.

\section{The modular invariance of the 4-particle amplitude}

First we note that the measure factor doesn't depend on the
specific choice of one of the zeroes of $Q=P_{2g-2}$. This can be
easily seen by noting the following relations between
$\omega_{PQ}$:
\begin{equation}
\omega_{P_aQ}(z) = \omega_{P_a\tilde Q}(z) -  \omega_{ Q \tilde Q
}(z).
\end{equation}
This induces a linear changes of basis of $\phi^a$ if we change to
another zero point of $\tilde Q\in\{P_1,\cdots, P_{2g-3}\}$ of
$\omega(z)$. So the measure factor $\wedge^a W^a$ just changes a
sign.

To prove the the modular invariance of the 4-particle amplitude
given in eq.~(\ref{conjecture}), let us consider a modular
transformation $\Gamma_g$:
\begin{equation}
\Gamma_g = \left( \begin{array}{cc} D & C\cr B & A \end{array}
\right)  \quad \in  Sp(2g,Z) .
\end{equation}
The holomorphic 1-forms $\omega_i$ and the period matrix $\Omega$
transform as follows:
\begin{eqnarray}
\omega_i & \to & \tilde\omega_i = \omega_j(C\Omega + D)^{-1}_{ji}
, \\
\Omega & \to & \tilde\Omega = (A\Omega + B)(C\Omega + D)^{-1} .
\end{eqnarray}
By using these results we have:
\begin{eqnarray}
\det\tilde\omega_i(z_j) & = &  \det(C\Omega + D)^{-1}
\det\omega_i(z_j), \\
\det {\rm Im}\tilde\Omega & = & |\det(C\Omega + D)|^{-2}\,  \det
{\rm Im}\tilde\Omega .
\end{eqnarray}
This proved that the following combination:
\begin{equation}
{|{\cal Y}_s|^2 \over (\det{\rm Im}\Omega)^4}, \end{equation} is
modular invariant $(g\ge3)$.

As $\omega$ doesn't depend on the choice of  homology cycles used
to defined the basis $\omega_i(z)$, one sees that $W^1\wedge W^2
\wedge\cdots W^{2g-3}$ is modular invariant. On the other hand $
W_i$ transforms identically as $\omega_i$:
\begin{equation}
W_{i} \to \tilde W_i = W_j (C\Omega + D)^{-1}_{ji} .
\end{equation}
This gives
\begin{equation}
\wedge_i W_i \to \wedge_i \tilde W_i = \det(C\Omega + D)^{-1} \,
\wedge_i W_i ,
\end{equation}
and so the following combination:
\begin{equation}
 { |\wedge^a W^a \wedge_i
W_i|^2 \over \det {\rm Im} \, \Omega  }, \end{equation} is also
modular invariant. This proves that the 4-particle amplitude given
in eq.~(\ref{conjecture}) is modular invariant.

\section{The factorization of the 4-particle amplitude}

Now we prove that the 4-particle amplitude also satisfies the
factorization condition.

To study the factorization, let us consider the dividing limit of
the Riemann surface\footnote{We will not consider the other limit
of pinching a non-zero homology cycle which would give a
6-particle amplitude .}. One way of taking this limit is to
construct a family of degenerating surfaces near $D_{g_1}$ over
the unit disk $D=\{t\in C| |t|<1\}$ as follows
\cite{Fay,VerlindeNPB}. Take two surfaces $\Sigma_1$ and
$\Sigma_2$ of genus $g_1$ and $g_2=  g - g_1$. Choose on each
surface a point $p_i$, $i=1,2$, and a coordinate neighbourhood
$U_i = \{ z_i | |z_i|<1\}$ near each $p_i$, such that $p_i = \{
z_i = 0\}$. Remove a small disk $|z_i|<|t|^{1/2}$ from both
surfaces, and glue the remaining surfaces together by attaching
the annulus $A_t = \{ w | |t|^{1/2} <|w| <|t|^{-1/2}\}$ according
to
\begin{equation}
w = \left\{
\begin{array}{ll} {t^{1/2}\over z_1} & \hbox{if}~|t|^{1/2}<|w|<1,
\cr t^{-1/2}z_2 \qquad & \hbox{if}~1<|w|<|t|^{-1/2}.
\end{array}
\right. \end{equation}
We denote the resulting surface by
$\Sigma_t$. The two components of $\Sigma_t-A_t$ we call $X_t$ and
$Y_t$ respectively. The parameter $t$ (which should not be
confused with the standard Mandelstam variable $t=-(k_2+k_3)^2$),
the points $p_1$ and $p_2$, together with the moduli of $\Sigma_1$
and $\Sigma_2$ provide a parametrization of $M_g$ near $D_{g_1}$.
Furthermore $t$ is the correct analytical coordinate on $M_g$
transversal to $D_{g_1}$ for $g_1>1$. (Near $D_1$ the analytic
transversal coordinate is $t^2$, as one can see explicitly in
\cite{XiaoZhu} by explicit computation. This is because the
punctured surfaces at $D_1$ have an automorphism of order two,
which means that $t$ and $-t$ correspond to the same points on
$M_g$.) The measure near the boundary ($t \to 0$) behaves as:
\begin{equation}
\wedge^aW_a \wedge_i W_i \to \wedge^aW_a^{(1)} \wedge_i W_i^{(1)}
\wedge {\rm d} p_1 \wedge {\rm d} p_2 \wedge {\rm d} t
\wedge^aW_a^{(2)} \wedge_i W_i^{(2)} .
\end{equation}
Because all determinants cancel, there is no singular factor of
$t^{-2}$ which appears in bosonic string theory
\cite{Knizhnika,GavaIengo,FS}.

Choosing a generic abelian holomorphic differential $\omega(z,t)$
on $\Sigma_t$. At leading order in $t$, this will have $2g_1-2$
zeroes on $X_t$ and $2 g_2 -2$ zeroes on $Y_t$. There are two more
zeros on $A_t$ which we denoted as $Q_1$ and $Q_2$. In the limit
$t\to 0$, $Q_i \to p_i$. The basic relation is:
\begin{equation}
\Delta_g(z_1,z_2)   =  O(t), \end{equation} if both $z_{1,2} \in
X_t$, or both $z_{1,2}\in Y_t$, and
\begin{eqnarray}
\Delta_g(z_1,z_2)   & = &  \pm   \Delta_{g_1}(z_1,Q_1)
\Delta_{g_2}(Q_2,z_2) + O(t)  \nonumber \\
& = & \pm \Delta_{g_1}(z_1,p_1) \Delta_{g_2}(p_2,z_2) + O(t),
\end{eqnarray}
if $z_{1} \in X_t$ and $z_2 \in Y_t$. We will not present the
proof of the above results here.

By using these results we have
\begin{eqnarray}
{\cal Y}_s  &  \to &  \pm (t+u-s)
\Delta_{g_1}(z_1,p_1)\Delta_{g_2}(p_2,z_3)
\Delta_{g_1}(z_1,p_1)\Delta_{g_2}(p_2,z_4) \nonumber \\
&  =  & \mp 3 s\, {\cal Y}_s^{(3)}(z_1,z_2,p_1) {\cal
Y}_s^{(3)}(z_3,z_4,p_2),
\end{eqnarray}
which shows that ${\cal Y}_s$ factorizes holomorphically into a
product of two 3-particle integrands.

Under the dividing degeneration limit, the exponential factor
approaches the following limit ($\alpha' = 2$):
\begin{eqnarray}
& & |t|^{ (k_1+k_2)^2} \, \langle {\rm e}^{ i k_1\cdot X(z_1)}
{\rm e}^{ i k_2\cdot X(z_2)} {\rm e}^{ -i (k_1 + k_2) \cdot
X(p_1)}
\rangle \nonumber \\
& & \qquad \times \langle  {\rm e}^{ i k_3\cdot X(z_3)}  {\rm e}^{
i k_4\cdot X(z_4)} {\rm e}^{  i (k_1 + k_2) \cdot X(p_2)} \rangle
.
\end{eqnarray}
Integrating over $t$ gives exactly the (massive) propagator:
\begin{equation}
\int {\rm d}^2  t |t |^{ (k_1+k_2)^2} \sim { -2 \pi \over s - 2} .
\end{equation}
This finishes the proof that the 4-particle amplitude of
eq.~(\ref{conjecture}) also satisfies factorization condition. Of
course the above analysis should be modified for $g_1 = 1$.

\section*{Acknowledgments}

 I would like to thank Edi Gava, Roberto Iengo and K. S. Narain for
helpful discussions, reading the paper and comments. Their
interests in this work greatly speed up the progress of my search
for the elusive multi-loop amplitudes.  Thanks also go to Jun-Bao
Wu and Zhi-Guang Xiao for their initial participation in this
work. I would also like to thank Bo-Yu Hou and CCAST (China Center
for Advanced Science and Technology) for the invitation to
participate the Workshop on the Integrability of Quantum Gauge
Field and Classical Supergravity from Jan. 12 to 22, 2005. This
work is supported in part by funds from National Natural Science
Foundation of China with grant number 10475104. Finally the author
would also like to thank Prof. S. Randjbar-Daemi  and the
hospitality at Abdus Salam International Centre for Theoretical
Physics, Trieste, Italy.

\end{document}